\documentclass[]{interact}

\usepackage{epstopdf}
\usepackage{easyReview} 
\usepackage{caption}
\usepackage{subcaption}

\usepackage{natbib}
\bibpunct[, ]{(}{)}{;}{a}{}{,}
\renewcommand\bibfont{\fontsize{10}{12}\selectfont}

\theoremstyle{plain}

\theoremstyle{definition}

\theoremstyle{remark}

\usepackage[hyphens]{url}
\usepackage{amsmath,amssymb,amsthm} 

\usepackage{import}

\begin{document}


\title{Observation of wave propagation over 1,000 km into Antarctica winter pack ice}

\author{
\name{Takehiko Nose\textsuperscript{a}\thanks{CONTACT Takehiko Nose. Email: tak.nose@edu.k.u-tokyo.ac.jp}, Tomotaka Katsuno\textsuperscript{a}, Takuji Waseda\textsuperscript{a}, Shuki Ushio\textsuperscript{b,c}, Jean Rabault\textsuperscript{d}, Tsubasa Kodaira\textsuperscript{a}, and Joey Voermans\textsuperscript{e}}
\affil{\textsuperscript{a}Department of Ocean Technology, Policy and Environment, Graduate School of Frontier Sciences, The University of Tokyo, Kashiwa, Chiba, Japan;\newline
\textsuperscript{b}National Institute of Polar Research, Tachikawa, Tokyo, Japan,\newline
\textsuperscript{c}The Graduate University for Advanced Studies, SOKENDAI, Tachikawa Tokyo, Japan,\newline
\textsuperscript{d}Norwegian Meteorological Institute, IT Department, Blindern N – 0313 OSLO, Norway;\newline
\textsuperscript{e}The University of Melbourne, Parkville VIC, Australia}
}

\maketitle
\begin{abstract}
A drifting wave-ice buoy, which was configured by mounting the OpenMetBuoy on an ad hoc floating platform that we named Medusa, was deployed at the Lützow-Holm Bay (LHB) marginal ice zone in Antarctica on 4 Feb 2022 during the 63rd Japanese Antarctica research expedition. The wave-ice buoy, Medusa-766, survived the Antarctica winter as the measurement duration reached 333 days.  
During the winter months, it was located deep in the ice cover with the shortest distance to the ice-free Southern Ocean over 1,000 km; at this time, there was evidence of ocean wave signals at the buoy position. Using the directional wave spectra obtained from the ECMWF's reanalysis, we show that the Medusa-766 observed waves were likely generated by an extratropical cyclone in the Southern Ocean. Wave-induced ice breakup potential for such an event could extend 100s km into the ice field. When Medusa-766 was in LHB in the summer months, it did not detect sizeable wave energy despite the low sea ice concentration extent even during on-ice waves events. Characterising the considerable differences in the wave attenuation at LHB is needed to elucidate the relative contribution of ocean waves to the unstable LHB fast ice. The success of Medusa-766 demonstrates the robustness of the general design, hardware, firmware, and the high sensitivity of the sensor used.  The result is promising for future LHB wave-ice interaction research. 
\end{abstract}

\begin{abbreviations}
	AMSR2 - Advanced Microwave Scanning Radiometer 2;
	ADS - The Arctic Data archive System;
	ERA5 - ECMWF Reanalysis v5;
	GNSS - Global Navigation Satellite System;
	IMU - Inertial measurement unit;
	JARE - Japanese Antarctica research expedition;
	LHB - Lützow-Holm Bay;
	MSLP - mean sea level pressure
	MIZ  - Marginal ice zone;
	MCU - Micro controller unit;
	OMB - OpenMetBuoy;
	SIC - Sea ice concentration;
	SAR - Synthetic aperture radar;
	PSD - power spectral density;
\end{abbreviations}


\section{Introduction}\label{intro}
The inaugural JARE was conducted in 1956; each year, a Japanese icebreaker traverses the summer ice cover to deliver necessary supplies to the Syowa Station. The current icebreaker \textit{Shirase} has been sailing since 2009. Variety of monitoring and research observations are made including atmosphere, ocean, and space. There are several anecdotal evidence that waves traverse the sea ice to reach the Syowa Station. 
One of the most striking event was the fast ice breakup that occurred on 18 Mar 1980, which led to the loss of two aeroplanes moored on ice near the Syowa Station. This event is documented in \citet{Higashi1982}, in which they concluded that the catastrophic ice breakup event near the Ongul Island was caused by the swell-induced flexural failure of the fast ice. An extratropical cyclone with a minimum pressure of 940 hPa generated waves that propagated towards the Syowa Station. \citet{Higashi1982} estimated the incoming wave height at the ice edge to be around 4 m, which was attenuated by sea ice cover to 0.4 m over a distance of 10 km. They suggested that such wave conditions  can disintegrate 1 m thick ice plate into ice pieces of less 100 m. 
The ice in this area was used as a runway for aeroplanes since 1957 when the Syowa Station was established, and the 1980 event was the first time fast ice broke up on a regional scale. 

\citet{Higashi1982} noted that even before the 1980 breakup event, fast ice extensively drifted out of LHB several times after severe storms. The uniqueness of the JAREs is the availability of decades of observations such as in situ Syowa Station measurements and navigation logs at almost the same shipping route and season. Exploiting this, \citet{Ushio2003,Ushio2004,Ushio2006} investigated the frequency of fast ice breakup in and its subsequent loss from LHB. They combined the in situ and navigation data with the satellite imagery and focused on the period between 1980 and 2004. 
Figure 2 in \citet{Ushio2006} shows that fast ice breakups occurred every year from 1980 to 1988. A stable period followed between 1988 to 1996, but after that, prolonged, e.g., 2--3 months, breakups occurred every year until the end of the analysis period in 2004. Satellite imagery shows that fast ice in LHB has been unstable with repeated breakups followed by a stable condition with wide area freezing up every several years.  Here, the "breakup" is defined as the offshore drift of fractured ice floes northward of 68$^\circ$ 50' S \citep{Ushio2006}. \citet{Ushio2003,Ushio2006} explored factors that contribute to the fast ice breakup and concluded that low snow cover over fast ice and the pack ice conditions northward of the LHB fast ice correlated with the LHB ice breakups. Regarding the latter point, \citet{Ushio2006} hypothesised that a pack ice zone serves as a "barrier" to the LHB fast ice. He suggests that a compact pack ice zone is a more effective barrier and showed that there is coincidence of LHB fast ice breakups and the formation of a large polynya in the pack ice zone during the winter months. To further consolidate this hypothesis, we need to quantify how much wave energy remains unattenuated and propagates to the LHB fast ice.

With the viewpoint to study the relative contribution of ocean waves to the LHB fast ice breakup, we conducted a wave measurement experiment during JARE63 and deployed four drifting wave-ice buoys and two SOFAR Spotter buoys in the MIZ offshore LHB.
Two of the four wave-ice buoys (a wave-ice buoy is a term used when a wave sensor is placed on an ice floe, and that ice floe serves as the sensor's floating platform) were equipped with the OMB \citep{Rabault2022} as the wave measuring sensor, of which one, named Medusa-766, measured and transmitted GNSS and wave data for around 333 days. Because Medusa-766 survived the harsh Antarctica winter, it recorded valuable observation deep in the winter pack ice: the data provide remarkable evidence that ocean wave energy propagates over 1,000 km into the ice cover. We use the well-known wave attenuation law to examine the likely source of the wave signal. We, then, discuss the distance scale of wave-induced ice breakup potential along the propagation distance. Wave-induced ice breakup is not well understood because of the difficulty in conducting observations that capture the exact instance of ice breakups. Despite the challenges, \cite{Voermans2020} proposed a wave-induced ice breakup index known as $I_{br}$ based on wave conditions and sea ice properties. We discuss the derivation and application of this wave-induced breakup index.
Lastly, the wave data measured in LHB is discussed in the context of wave attenuation characteristics.

\section{Wave-ice buoy, data, and analysis methods} \label{JARE63overview}
\subsection{The Medusa-766 wave-ice buoy}\label{buoy}
A reliable and durable wave-ice buoy is desired for successful deployments in the harsh Antarctica sea ice cover. Additionally, long-lasting battery life is another desired trait as the opportunity to deploy sensors in Antarctica are limited. To this end, the OMB \citep{Rabault2022} was considered a suitable choice for the JARE wave-ice observation for the following reasons. First, the OMB uses a low power MCU and an industrial grade IMU that is thermally calibrated to $-40$ \textcelsius  \citep{STMicro}. While microprocessor based wave-ice buoys have been used to collect wave-ice data, e.g., \citet{Kohout2015,Rabault2020}, low power MCUs, which have sufficient clock speed and RAM to perform necessary onboard calculations, have advanced the wave-ice buoy technology. The OMB reliability has been proven in the Arctic region \citep{Rabault2023,Nose2023}, while the JARE63 observation was the first OMB deployment on the Antarctica sea ice.   

We outfitted the OMB with six Tadiran Lithium D-cell TL-5930 batteries each with 19 Amp-hr capacity. The number was six, not because we planned a specific measurement duration, but rather we packed as many batteries as we can into the sensor enclosure (the Takachi waterproof enclosure SPCP131815T). From previous campaigns, we expected battery life of one year. A detailed description of the estimated power consumption during the observation period is given in Appendix \ref{Iridium_analysis}. The sensor enclosure was mounted on a substructure, or a floating platform, that was designed for the JARE63 deployment.
In many ice floe observations, a sensor and its enclosure are directly placed on the ice floe. However, in severe ice conditions of Antarctica winter, the sensor enclosures can be destroyed by ridging and rafting of the ice. The ad hoc floating platform that we named Medusa was designed to provide sensor protection, on which the OMB was mounted. 
Medusa was 10 kg in weight and its dimensions were 540 mm diameter and 280 mm in height. Medusa's structure consists of a life-saving buoy that is sandwiched with two stainless steel plates, on which the sensor enclosure is mounted. 
Figure \ref{fig3:Medusa766} presents Medusa-766 images: the left image shows how the sensor enclosure was attached to Medusa and the right image shows the Medusa-766 being deployed on an ice floe. 

Spectral noise in accelerometer-based wave buoys affects wave measurements, e.g., overestimation of spectral density in the frequency range 0.05--0.15 Hz when wind speed exceeds 10 ms$^{-1}$ or wave heights above 4 m \citep{Collins2022}.  With the emergence of IMU-based wave buoys housed in a relatively small platform (diameter$\approx$0.5 m or smaller), we seem to be facing a similar but different problem regarding the noise floor. When the wave-ice buoy motion is stable, e.g., on an ice floe with diameter $>$ encountering wave wavelengths, the spectral noise floor is consistent with the sensor specification white noise; however, when the buoy floats in open ocean, e.g., our Arctic Ocean wave observations \citep{Waseda2017,Nose2018,Nose2023}, the spectral noise floor significantly elevates and the ideal filter is needed to retrieve the frequency integrated wave statistics. As will be shown later, the results discussed in this study is unaffected by the spectral noise floor.

The default GNSS and wave measurement intervals were 0.5 and 1 hour, respectively. The 1 hourly wave intervals were chosen as we anticipated frequent detectable wave energy during the low ice extent months between February and April. After April, we changed the measurement intervals to 1 and 3 hours (GNSS and wave), exploiting the 2-way Iridium messaging capability \citep{OMBGit} that was implemented for the JARE63 campaign. In the harsh field environment, however, the OMB MCU periodically reboots when firmware logic somehow goes amiss; then, the measurement intervals return to its default intervals. After a reboot, we need to re-send another message to change the intervals back to the desired duration.

\subsection{Atmospheric pressure and wind, sea ice, and wave data}
The MSLP and wind data were obtained from ERA5 and used to describe the synoptic conditions in the Southern Ocean. The SIC was used to infer Medusa-766 surface conditions and to estimate the distance from the ice edge to its position when there was wave signal. The SICs are derived based on the AMSR2 data and obtained from ADS \citep{ADS-AMSR2} (ADS-AMSR2 herein). 

Regional wave fields were produced from significant wave height, peak wave period, and mean wave direction from ERA5, which was used to describe the wave field evolution under extratropical cyclones and the wave conditions near the ice edge. The directional wave spectra $S(f,\theta)$ were obtained also from ERA5 and used to interpret the incoming wave directional distribution that propagates into the ice field. $f$ is the frequency (Hz) and $\theta$ is the wave direction in the meteorological convention (i.e., waves propagate from). The $S(f,\theta)$ integrated over the directions produce the frequency wave spectra (or PSD) as $S(f)=\int S(f,\theta)d\theta$, which was used to estimate the wave attenuation rate due to sea ice. Instead of integration over all directions, limiting the integration range can produce discretised directional energy that propagates towards the buoy position. 

\subsection{Wave statistics and wave-ice attenuation}\label{analysis_method}
Wave spectra (PSD) of the vertical surface displacements are estimated from the IMU readings and transmitted via Iridium messages \citep{Rabault2022}. The integrated wave statistics are calculated from the $S(f)$. The significant wave height is $H_{m0}= 4\sqrt{m_0}$ where $m_0 = \int_{f0}^{f1} S(f)df $. Wave periods used were the peak period $T_p$, which is the inverse frequency of the peak $S(f)$, and the $-1$ moment period, also known as the energy mean wave period in ERA5, is $T_{0m1}=\frac{\int_{f0}^{f1} f^{-1}S(f)df}{m_0}$. The frequency range [$f0,f1$] was [0.044,0.503] Hz for Medusa-766 and [0.034,0.548] Hz for ERA5.

To analyse how much wave energy attenuated from the incoming waves, we used the well-known form of wave attenuation by ice, 
\begin{equation}\label{eq3}
S_{ice}(f) = S_{in}(f)e^{-\alpha x},
\end{equation}
where $S_{ice}$ and $S_{in}$ are the attenuated and incoming PSDs, and $x$ is a distance between the two points. 
The wave attenuation is frequency dependent via
\begin{equation}\label{eq4}
	\alpha = af^n,
\end{equation} which was first observed by \citet{Wadhams1975}. 
The constant $a$ and the exponent $n$ are understood to vary depending on ice types \citep{Meylan2018,Voermans2021}. 

\section{Results: ocean wave signal over 1,000 km into ice cover} \label{wave_analysis}
\begin{figure}[h]
	\centering
	\includegraphics[width=\textwidth]{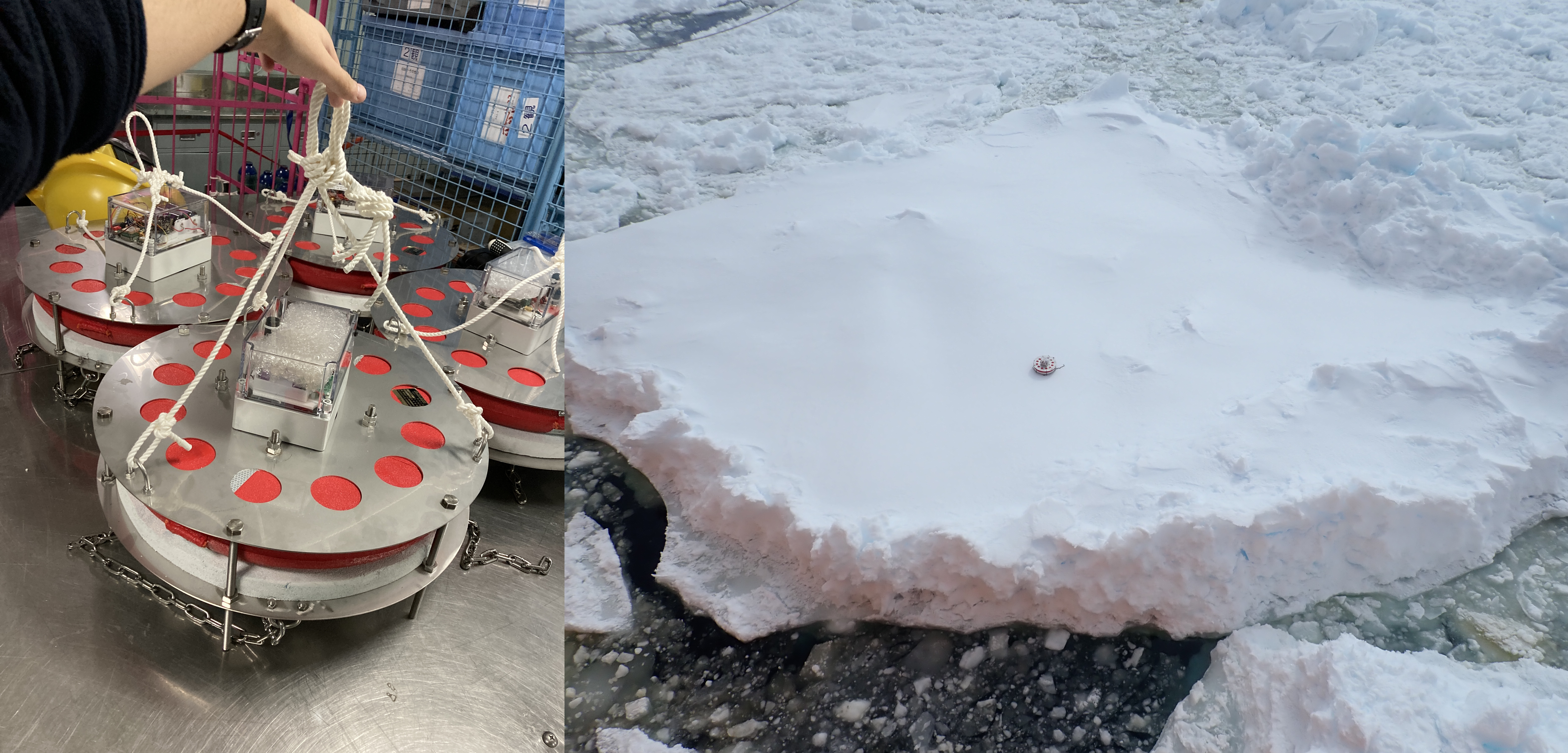}
	\caption{Images of Medusa-766. The left picture shows how the sensor box was attached to Medusa. The right picture shows Medusa-766 on the ice floe. Medusa's diameter is 540 mm, and the floe size was estimated to be around 20 m in the field. The photos were taken by T. Katsuno.}
	\label{fig3:Medusa766}
\end{figure}

\begin{figure}[h]
	\centering
	\includegraphics[width=\textwidth]{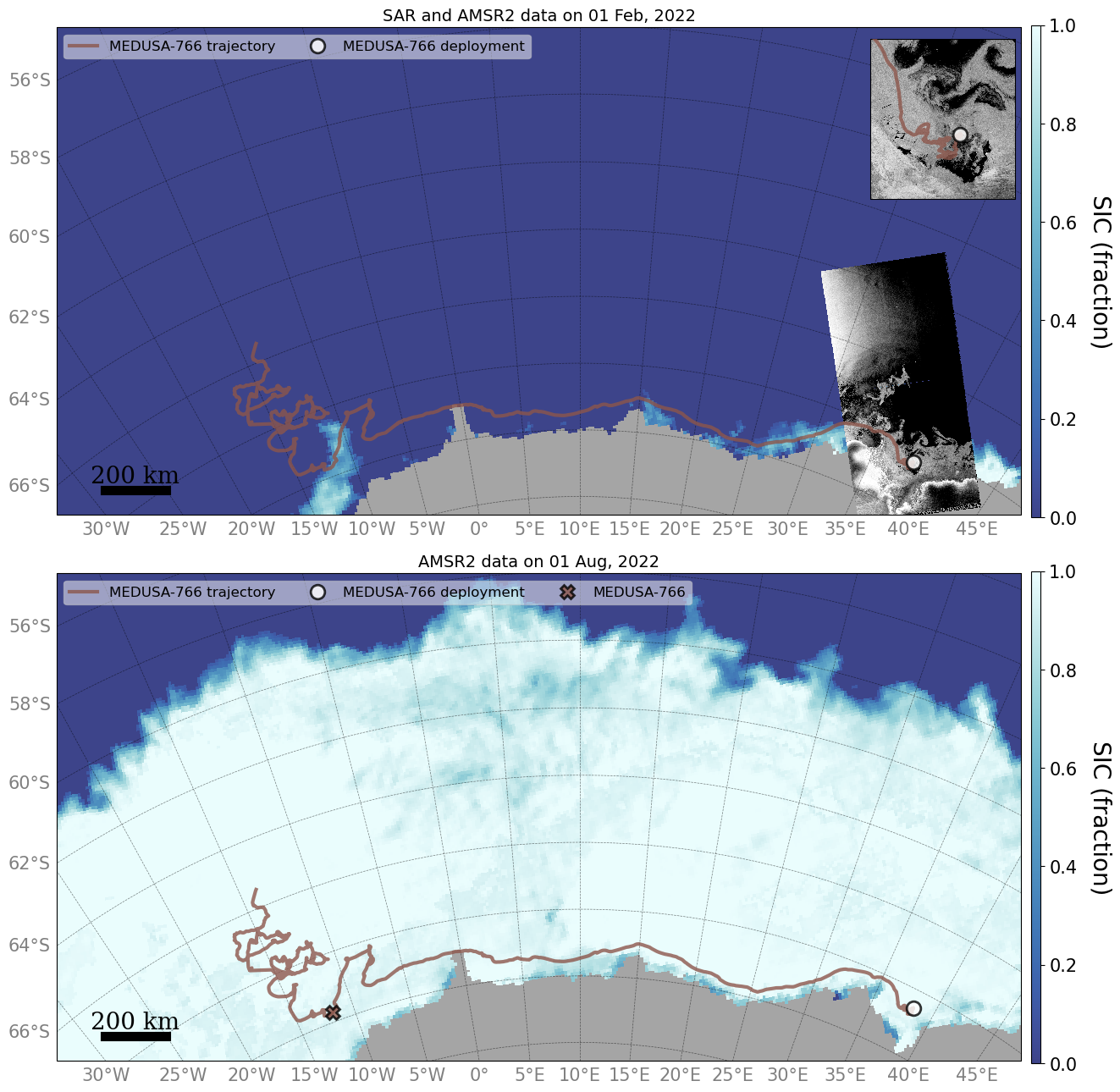}
	\caption{Medusa-766 trajectory between 4 Feb 2022 and 3 Jan 2023 overlaid on the ADS-AMSR2 SIC on 1 Feb (top) and 1 Aug (bottom) 2022. The brown line is the trajectory, and the brown marker shows its location on the respective dates. SICs are shown in colours. The 1 Feb panel also has Sentinel-1 SAR images overlaid.}
	\label{fig4:trajectory}
\end{figure}

Medusa-766 was deployed at 68.54$^\circ$ S, 38.29$^\circ$ E on 4 Feb 2022. It was placed on an ice floe of roughly 20 m wide by a crane from \textit{Shirase} (see Figure \ref{fig3:Medusa766}). The last contact from Medusa-766 was on 3 Jan 2023. The total measurement duration was approximately 333 days, and the rates of data acquisition and transmission success for both the GNSS positions and wave spectra were \textasciitilde95 \% (details of the Iridium transmission results are provided in Appendix \ref{Iridium_analysis}). It recorded over 10,000 GNSS positions and 4,000 wave spectra, and it drifted a total estimated distance of over 5,000 km. The Medusa-766 trajectory is plotted in Figure \ref{fig4:trajectory} with the ADS-AMSR2 SIC fields on 1 Feb and 1 Aug 2022. The figure shows the vast scale of seasonal sea ice cover variability in Antarctica during 2022. 
Here, we present the wave signal measured in the winter Antarctica pack ice when Medusa-766 was located over 1,000 km from the ice edge. 

\begin{figure}[h]
	\centering
	\includegraphics[width=\textwidth]{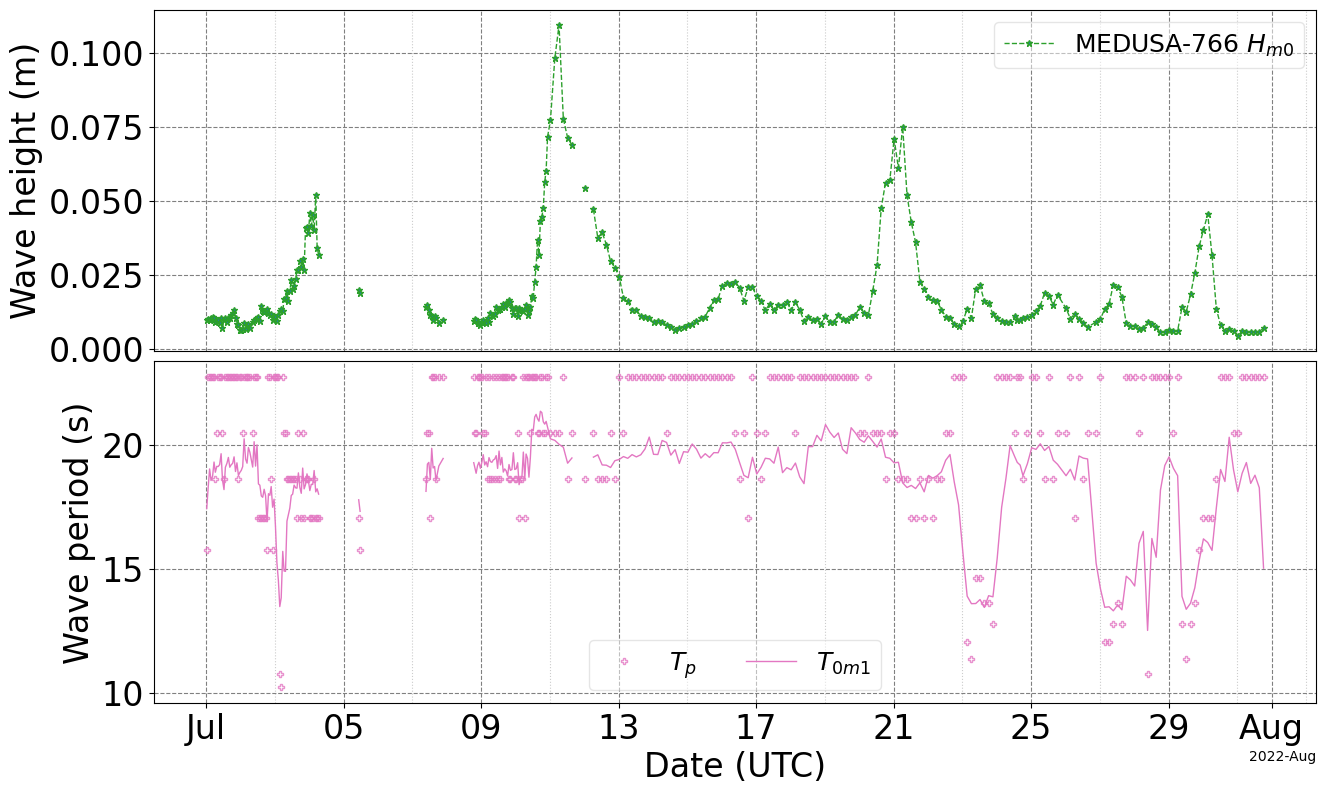}
	\caption{Medusa-766 wave height and period time series for July 2022. Medusa-766 was located deep in the Antarctica winter pack ice where the shortest distance to the ice-free Southern Ocean was over 1,000 km.}
	\label{fig2p5_JulTS}
\end{figure}

\begin{figure}[h]
	\centering
	\includegraphics[width=\textwidth]{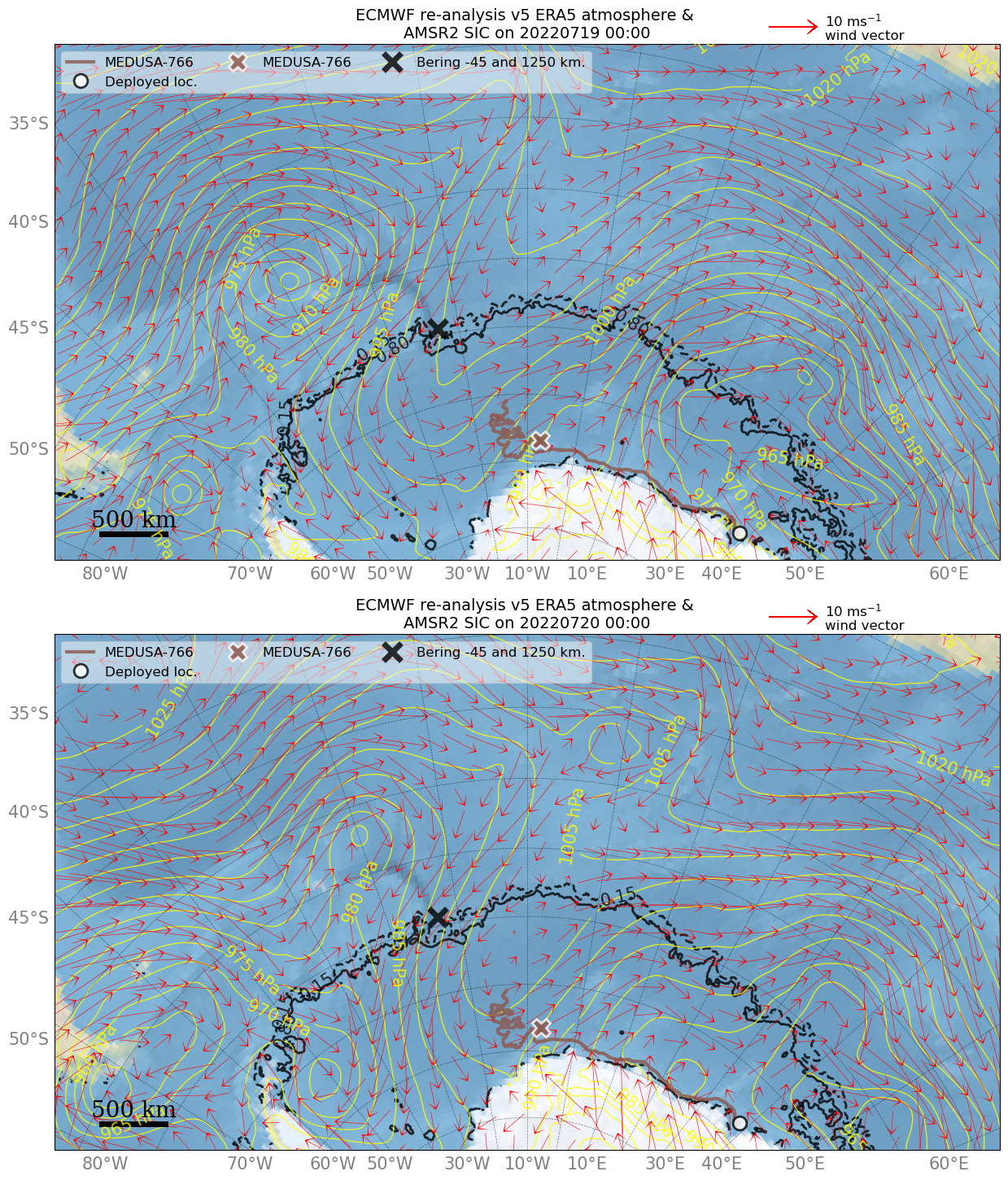}
	\caption{Synoptic conditions and the Medusa-766 positions for 19 Jul and 20 Jul 2022 (top and bottom, respectively) leading up to the wave event at the Medusa-766 position on 21 Jul. The brown line is the Medusa-766 trajectory, and the brown marker shows its location on the respective dates. The black marker is the location where the ERA5 directional spectra were obtained. ERA5 MSLP is shown in yellow contours, and the red vectors are the ERA5 10 m winds. The 0.15 and 0.80 ADS-AMSR2 SIC contours are also shown as black dashed and solid lines, respectively.}
	\label{fig3:mslp}
\end{figure}

\begin{figure}[h]
	\centering
	\begin{subfigure}{0.8\textwidth}
		\includegraphics[width=\textwidth]{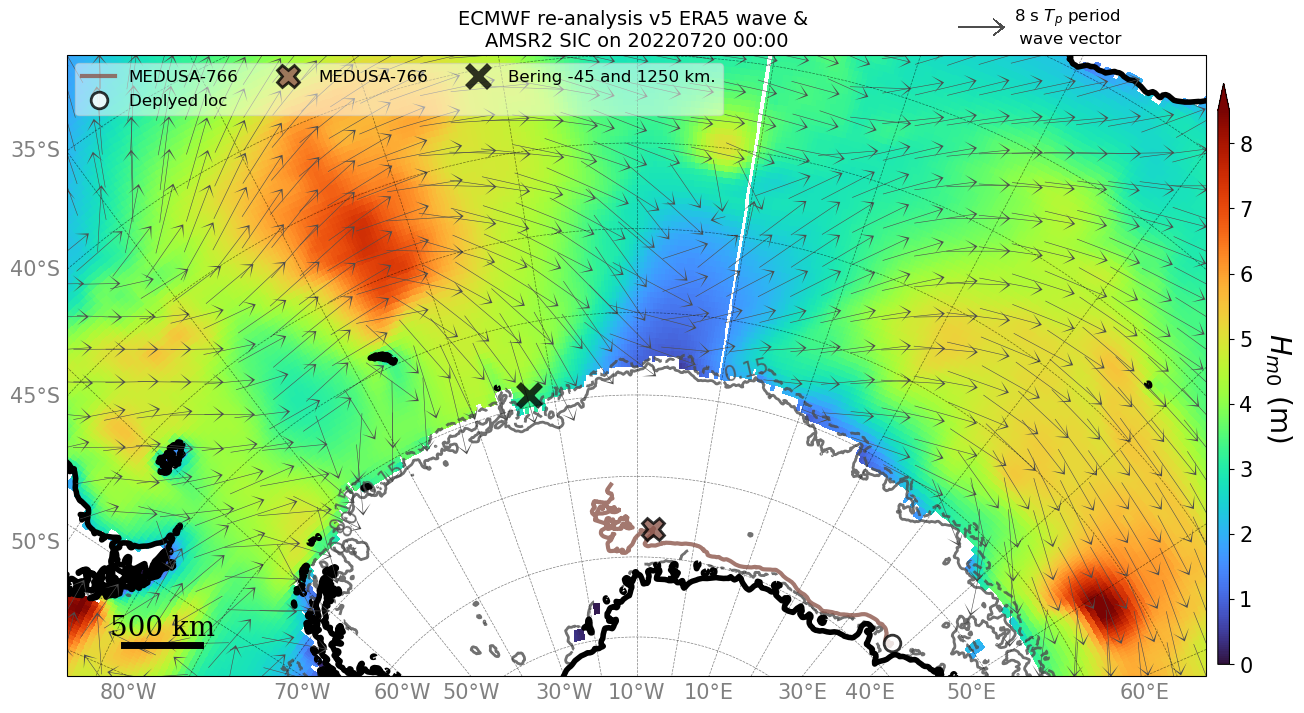}
		\caption{ERA5 wave field when Medusa-766 measured wave signal approximately 1,250 km into the Antarctica winter pack ice. The brown line is the Medusa-766 trajectory, and the brown marker shows its location at this time. The $H_{m0}$ is shown in the colours and wave vectors are mean wave directions scaled by $T_p$. The 0.15 and 0.80 ADS-AMSR2 SIC contours are also shown as grey dashed and solid lines, respectively. }
		\label{fig4a}		
	\end{subfigure}
	\begin{subfigure}{0.8\textwidth}
	\includegraphics[width=\textwidth]{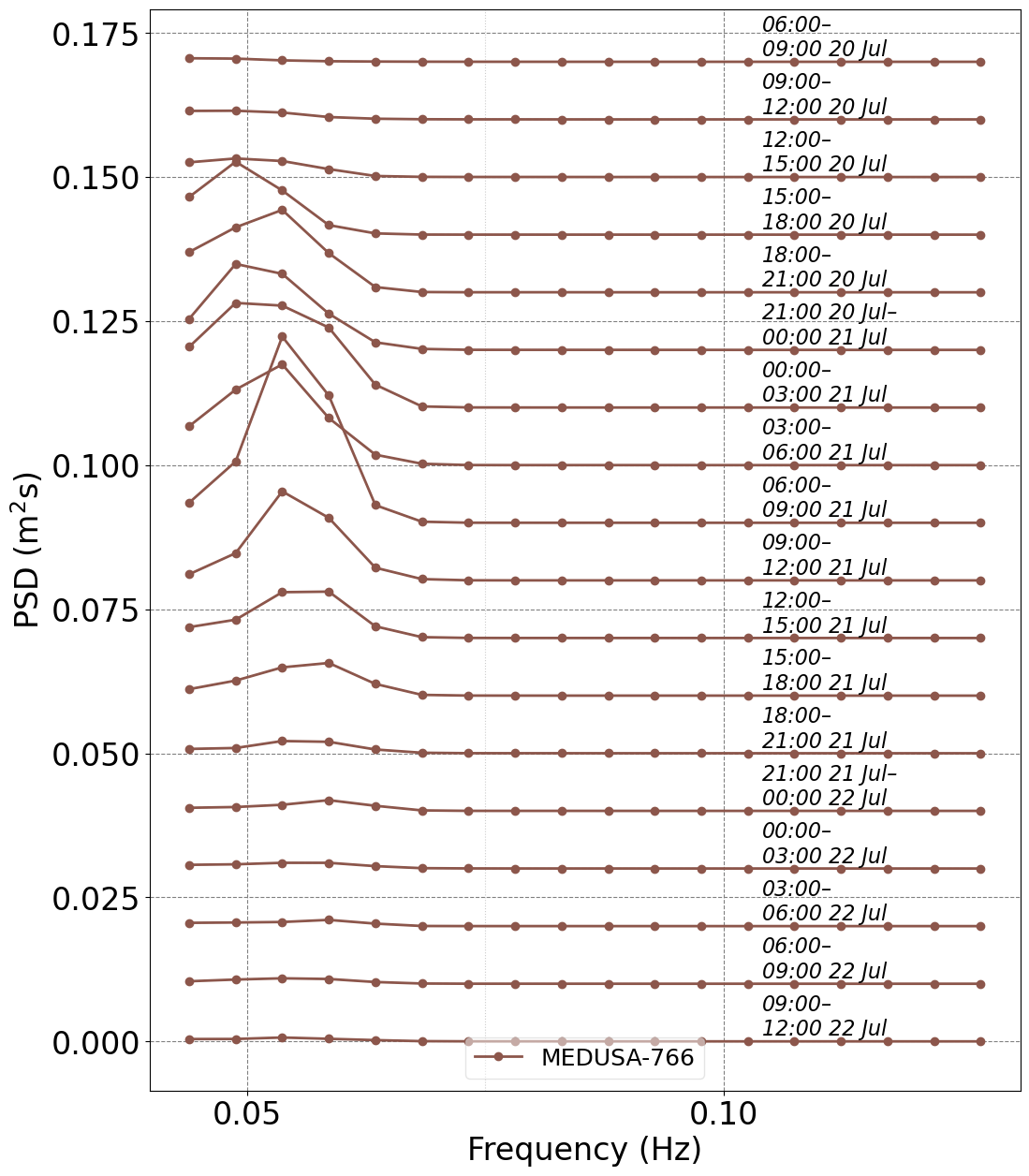}
	\caption{The wave evolution in the ice is shown in the PSD waterfall plot on the right panel; the wave signal peaked at UTC 06:20 on 20 Jul 2022.}
	\label{fig4b}
	\end{subfigure}	
	\caption{The wave field and Medusa-766 measured spectra on 20--21 Jul 2022.}
	\label{fig6:jul_event}
\end{figure}

By 10 Jul 2022, the ice cover between the Weddell Sea and LHB (60$^\circ$ W and 40$^\circ$ E) had advanced to around 60$^\circ$ S. This meant that Medusa-766 was deep in pack ice where the closest ice edge was over 1,000 km away. Because Medusa-766 has a low spectral noise floor of 60 $\frac{\mu g}{\sqrt{\textrm{Hz}}}$ \citep{STMicro}, which has been achieved in our observations, it is in theory capable of measuring centimetre order wave heights. 
During July and August, there were several events when we can identify ocean wave signal in the wave data based on the visual inspection of the spectra, in which the measured signal appears sufficiently higher than the noise level. 
We counted that Medusa-766 detected ocean wave signal deep in the pack ice four times in July and twice in August. The wave height and period time series measurd by Medusa-766 during July 2022 are presented in Figure \ref{fig2p5_JulTS}, which shows that the $H_{m0}$ values reached up to 10 cm. We focus on the event on 20--21 July. Leading up to this event, energetic waves were generated in the ice-free Southern Ocean off the Weddell Sea when the extratropical cyclone developed. This system appear to have developed near 47$^\circ$ S, 40$^\circ$ W on 00:00 18 Jul 2022 then migrated south to 55$^\circ$ S by 19 July with a minimum MSLP of 960 hPa. The system weakened to around 975 hPa as it migrated east to 30$^\circ$ W, and this was inferred as the probable synoptic condition that generated waves to Medusa-766. Figure \ref{fig3:mslp} depicts the described synoptic conditions on 19 and 20 Jul 2022. The wave field on 20 Jul 2022 is shown in Figure \ref{fig4a}, which indicates that the primary wave vectors at the ice edge are directed towards the pack ice zone.  The spectral evolution during the event is shown as a waterfall plot in Figure \ref{fig4b}.

\begin{figure}[h]
	\centering
	\includegraphics[width=\textwidth]{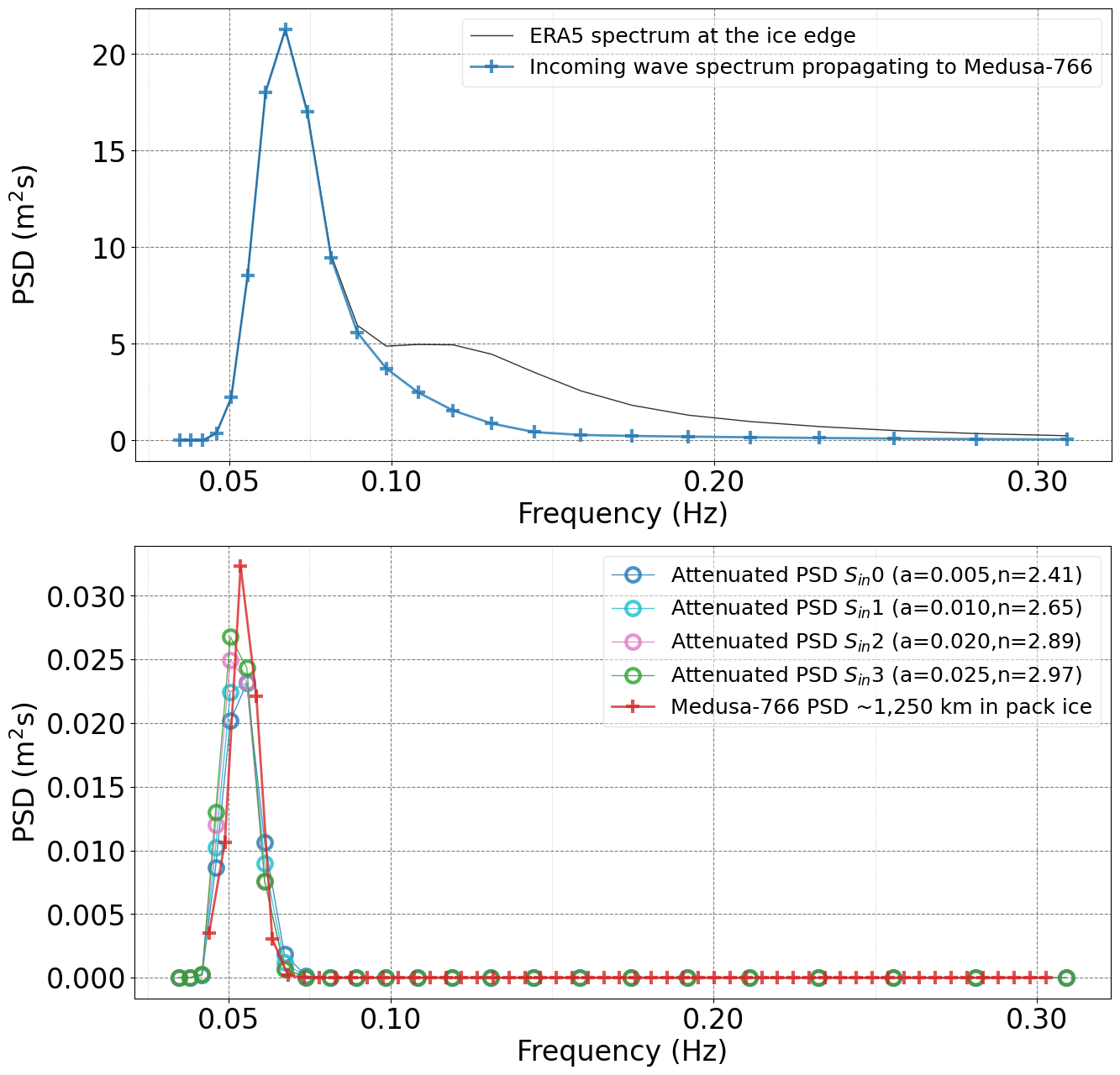}
	\caption{The ERA5 incoming wave spectrum at the ice edge is shown on the top panel. The black solid line was the PSD obtained by integration over all directions whereas the blue line with markers integrated the northwest sector. The ERA5 incoming wave, $S_{in}$ was attenuated using various combinations of $a$ and $n$ that matched the Medusa-766 $H_{m0}$ and shown on the bottom panel. A summary of the wave heights and periods from these PSDs are provided in Table \ref{tab:attenuation_summary}. }
	\label{fig7:spectra_event}
\end{figure}

To investigate whether Medusa-766 measured signals were the Southern Ocean waves that propagated into the ice field, we fitted the attenuation of an incoming wave spectrum to that of the observed Medusa-766 spectrum on 06:00 21 Jul 2022, which was the peak of the event (see Figure \ref{fig4b}). The ice edge incoming wave spectrum was obtained from ERA5 at 59$^\circ$ S, 22$^\circ$ W in the form of directional spectra (see the black marker in Figure \ref{fig4a}), and the distance between the incoming wave spectrum position and the Medusa-766 (68.2$^\circ$ S, 7.2$^\circ$ W) was approximately 1,250 km. At the incoming wave location, the $H_{m0}$ peaked at around 4 m with a $T_p$ of 14.9 s at UTC 09:00 on 20 Jul 2022. We note here that linear theory group speed of this wave system takes around 23 hours to travel 1,250 km. The directional spectrum revealed that there were two energy systems: one in the northwest and the other in the northeast sectors. The local wind vectors were directed from northeast, so we can assume the northeast waves were wind seas while the northwest waves were the swell, likely generated by the extratropical cyclone. The PSDs obtained from integrating the directional spectrum over all directions and discretised to the northwest sector are shown in the top panel of Figure \ref{fig7:spectra_event}. The primary wave system was indeed the swell energy propagating from the northwest sector towards the Medusa-766 position, and the directionally discretised northwest sector PSD was used as $S_{in}$ with $H_{m0}$ and $T_p$ of 3.25 m and 14.86 s.

The goal of this exercise is to show that Medusa-766 observed signal could be explained by attenuating the incoming wave spectrum using the well-known wave damping law (Equations \ref{eq3} and \ref{eq4}). It is debatable whether applying this attenuation form to describe the significantly attenuated signal over such a long distance is valid; indeed, we show here that many combinations of $a$ and $n$ may reproduce the observed wave statistics reasonably. We first refer to table 1 of \citet{Thomson2021} for typical values of constant $a$ in $\alpha \propto af^n$, which were (0.005,0.260). We selected four $a$ values 0.005, 0.010, 0.020, and 0.025, and tuned the exponent $n$ to attenuate the $S_{in}$ so that the observed Medusa-766 $H_{m0}$ was matched. The $n$ exponent values, tuned to two decimal places, were 2.41, 2.65, 2.89, and 2.97 respectively; $n$ is understood to range between 2 and 4 in the existing field studies \citep{Meylan2018,Thomson2021,Waseda2022} with an attenuation distance scale of $O(100)$ km. The Medusa-766 PSD and the attenuated PSD $S_{in}0-3$ are shown on the bottom panel of Figure \ref{fig7:spectra_event}. A summary of wave heights and periods, and the attenuation coefficients are given in Table \ref{tab:attenuation_summary}. The same exercise could be repeated to match the observed Medusa-766 $T_{0m1}$; however, due to the fact that many combinations of $a$ and $n$ exist, we are unable to attribute a physical meaning of the coefficients, e.g., inferring mechanism with which the wave energy is attenuated. Measuring unique attenuation coefficients is a motivation for future observations. Nevertheless, the similarity between the attenuated PSDs and the Medusa-766 observed PSD support that the captured signal was ocean waves regardless of the physical meaning of the attenuation coefficients.

\begin{figure}[h]
	\centering
	\includegraphics[width=\textwidth]{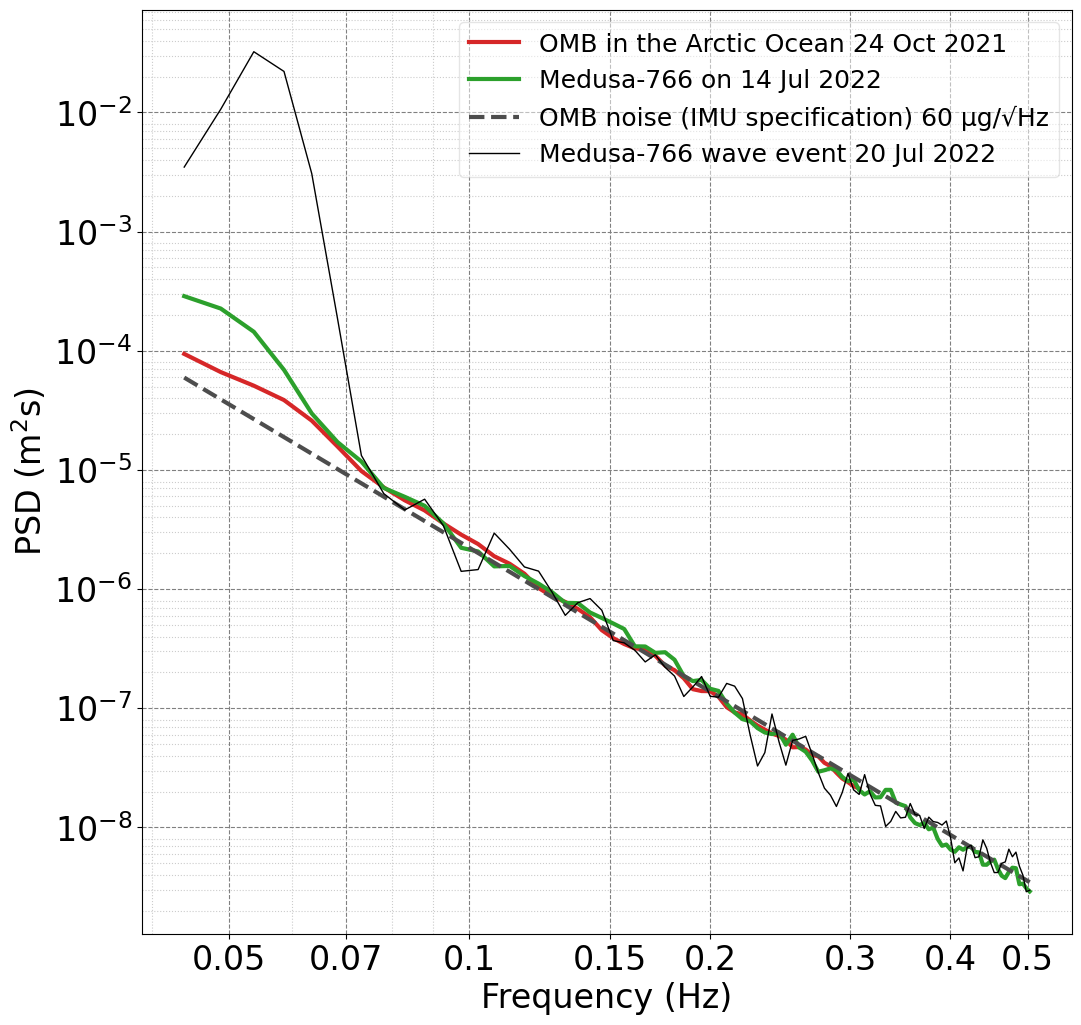}
	\caption{The OMB specification spectral noise floor and the field-achieved spectral noise floors from Medusa-766 and the Arctic Ocean observation \citep{Nose2023} are plotted with the Medusa-766 swell signal spectrum on 06:00 21 Jul 2022. The field-achieved noise floors were obtained by averaging spectra obtained on 14 Oct and 24 Jul 2022 for Medusa-766 and the Arctic Ocean observation, respectively.}
	\label{fig7:spectralnoise}
\end{figure}

We corroborate the 20--21 Jul swell event results by comparing the Medusa-766 spectrum with the noise floor achieved in the field (Medusa-766 itself as well as the Arctic Ocean observation \citep{Nose2023}) and the catalogue specification in Figure \ref{fig7:spectralnoise}. The daily averaged spectrum for 14 Jul 2022 for Medusa-766 and 24 Oct 2021 for the \citet{Nose2023} observation were considered to be the field-achieved noise floors. The catalogue specification IMU noise is $N_0 =60 \frac{\mu g}{\sqrt{\textrm{Hz}}}$, and the noise floor was estimated as $(N_0\times10^{-6})^2(2\pi f)^{-4}$. The observed swell energy is 2 orders of magnitude above the noise floors estimated from the field and the IMU specification. Figure \ref{fig7:spectralnoise} is a convincing demonstration of the sensor sensitivity and that the measured swell signal was unaffected by the sensor noise. For completeness, we list the respective ${H_{m0}}$ values: 0.4 cm for the catalogue specification, 0.5 cm for the Arctic Ocean noise floor, 0.8 cm for the Medusa-766 noise floor, and the 20--21 Jul event's $H_{m0}$ was 7.5 cm.

\section{Discussion}\label{discussion}
\subsection{Distance scale of wave-induced sea ice breakup}
With a viewpoint to gain insights into the wave-induced ice breakup potential, the ice breakup up index known as $I_{br}$ was calculated along the propagation distance based on the attenuation coefficients $a=0.01$ and $n=2.65$ in Equations \ref{eq3} and \ref{eq4}. 
Following \citet{Voermans2020}, the monochromatic wave breaks the ice when the largest stress imposed by a wave on an elastic ice sheet exceeds the flexural strength $\sigma_{flx}$ of the ice sheet: $\left(\frac{2\pi^2 A h_i}{\lambda_{mono}^2}\right)Y>\sigma_{flx}$ (e.g., \citet{Dumont2011}). This yields an ice breakup index for monochromatic waves,
\begin{equation}\label{eq1}
	I_{br}^{(mono)} = \left(\frac{2\pi^2 A h_i}{\lambda_{mono}^2}\right) \left(\frac{Y}{\sigma_{flx}}\right),
\end{equation}
in which the ice breaks up when $I_{br}^{(mono)}>1$. Here, $\sigma_{flx}$ is the ice flexural strength, $Y$ is the Young's modulus, $A$ is the monochromatic wave amplitude, $\lambda_{mono}$ is the wavelength, and $h_i$ is the ice thickness.
\citet{Boutin2018,Voermans2020} further extended this breakup index to incorporate ice breakups in a random wave field; the monochromatic wave amplitude and wavelength were replaced by the significant wave height $H_{m0}$ and peak wavelength $\lambda_p$. A coefficient was introduced to consider the ultimate limit state within a given time period as follows:
\begin{equation}\label{eq_c1}
	\frac{A}{\lambda_{mono}^2}\approx \frac{c_1 H_{m0}/2}{\lambda_p^2},
\end{equation}
where $c_1\frac{H_{m0}}{2}/\lambda_{p}$ may be considered as an approximation of the steepest expected wave in a given sea state. The breakup index for a random wave field is, then, derived as
\begin{equation}\label{eq2}
	I_{br}^{(rand)}	=\left(\frac{2\pi^2 c_1\frac{H_{m0}}{2} h_i}{\lambda_{p}^2} \right)\frac{Y}{\sigma_{flx}}.
\end{equation}
\citet{Voermans2020} omitted the constants ($c_1\times2\pi^2$) from the breakup index and found observation evidence that a threshold value is 0.014 adopting $c_1=3.6$ (proposed by \citet{Boutin2018} for a stationary wave field of 500 waves), which is equivalent to $I_{br}^{(rand)}=1$. The value of the coefficient $c_1$, however, remains debatable due to uncertainties associated with the flexural strength $\sigma_{flx}$ and the Young's modulus $Y$ \citep{Timco2010,Karulina2019}. Moreover, the underlying assumption of a brittle fracture without a plastic deformation remains to be tested. For now, we ignore the plastic deformation regime following \citet{Voermans2020}, and the $I_{br}^{(rand)}$ index was used to quantify the spatial scale of wave-induced ice breakup potential from observations and models.

The swell part of a spectrum has a coarse frequency resolution, which can cause discontinuities in $\lambda_p$ along the wave propagation distance (and in turn $I_{br}^{(rand)}$); as such, we used the wavelength of the $T_{0m1}$, denoted as $\lambda_{-1}$, instead of $\lambda_{p}$ in Equation \ref{eq2}. The $I_{br}^{rand}$ threshold is unchanged because the $c_1$ coefficient in Equation \ref{eq_c1} becomes 3.42 adopting $\lambda_{-1}\approxeq0.95\lambda_p$ (an applicable approximation for swell spectra as shown in the work of \cite{Ahn2021}). The $-1$ moment mean wavelength $\lambda_{-1}$ is estimated from the linear dispersion relation $\frac{g}{2\pi}T_{0m1}^2$ following \citet{Voermans2020}. Since ice properties were not measured, we use $\sigma_{flx}$ and $Y$ also from \citet{Voermans2020}: $\sigma_{flx} \in [0.1,0.7]$ MPa and $Y \in [1,6]$ GPa with most probable values of $\sigma_{flx}=0.4$ MPa and $Y=3$ GPa. 

\begin{figure}[h]
	\centering
	\includegraphics[width=\textwidth]{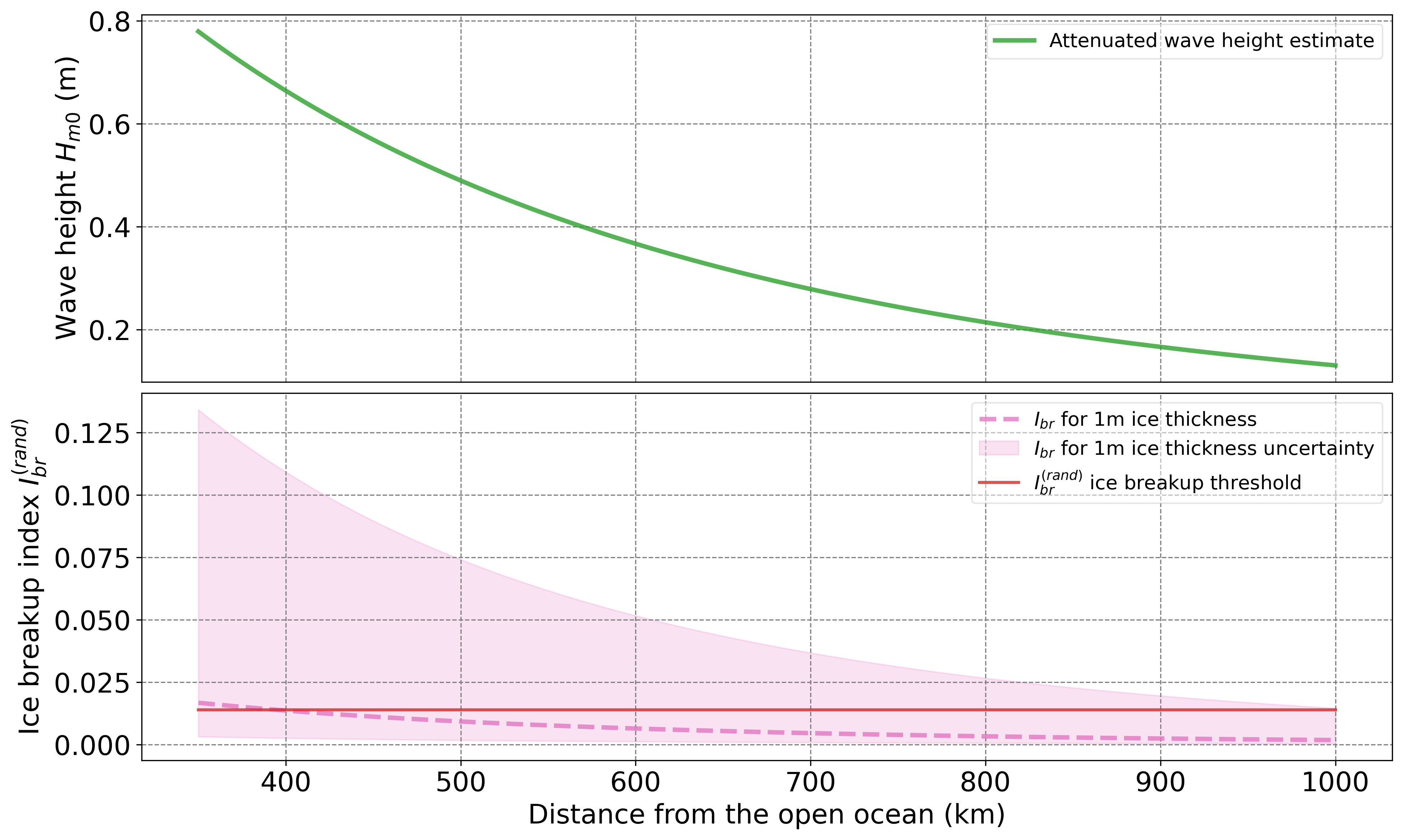}
	\caption{The attenuated wave height using Equation \ref{eq3} (top) is plotted with the wave-induced ice breakup index, $I_{br}^{(rand)}$, using Equation \ref{eq2} (bottom) along the wave propagation distance between 400 and 1,000 km from the ice edge. The uncertainty was calculated using $\sigma_{flx}=0.1$ MPa \& $Y=6$ GPa for the lower bound and $\sigma_{flx}=0.7$ MPa \& $Y=1$ GPa for the upper bound. }
	\label{figx_ibr}
\end{figure}

Assuming ice thickness of 1 m, and the probable ice properties ($\sigma_{flx}=0.4$ MPa and $Y=3$ GPa), the wave-induced ice breakup threshold of 0.014 was exceeded to around 400 km into the ice cover (see the bottom panel of Figure \ref{figx_ibr}). The scale here is difficult to comprehend; the reason is that considerable uncertainty arises from the lack of in situ sea ice mechanical properties $\sigma_{flx}$ and $Y$. As a demonstration, Figure \ref{figx_ibr} presents the attenuated wave estimates as waves travel into sea ice and the wave-induced ice breakup index $I_{br}^{(rand)}$ with a conservative uncertainty: the upper bound uncertainty using $\sigma_{flx}=0.7$ MPa and $Y=1$ GPa remains above the wave-induced ice breakup threshold up to 1,000 km into the ice cover.

The sea ice mechanical properties used in the $I_{br}^{(rand)}$ parameter is a considerable error source. In light of the analysis here, it is clear that more observations will improve our understanding of the ice breakup physics. To this end, a new approach to measure ice properties and the precise timing of ice breakups using geophones is emerging \citep{Moreau2020,Voermans2023}; these observations are promising for future ice breakup studies. 

\subsection{Waves in ice measurements in Lützow-Holm Bay: the JARE perspective}
From the winter Antarctica ice cover wave observation, we showed the ocean wave signal over 1,000 km into the pack ice likely originated from the extratropical cyclone in the Southern Ocean. This evidence seems robust, however, the analysis that followed needs to be tempered by the obvious fact that the interpretation was made from the single buoy measurements. For example, it is unlikely that the attenuation coefficients remain constant for the long propagation distance because we inherently assume the ice type remains the same too. Perhaps this is the reason that many combinations of the attenuation coefficients can achieve a tolerable fit to the observed $H_{m0}$. The wave-induced ice breakup potential for 1 m ice thickness could extend to 400 km from the ice edge, but if the ice breaks up, the wave attenuation characteristics are modified. The attenuation rate could be changing in time and space. Our lack of knowledge about the typical floe size along the propagation distance is another unknown. Lastly, many studies show that swell dispersion relation is practically unchanged between open water and typical ice field (e.g., figure 1 of \citet{Boutin2018}). However, there is a conspicuous observation of $\lambda$ shortening in \citet{Liu1988} who observed 18 s waves' $\lambda$ was around 250 m in the Antarctica pack ice whereas the linear dispersion relation derived $\lambda$ is over 600 m. Notwithstanding these limitations, the Medusa-766 observation shows that the waves likely propagate a strikingly long distance from the ice-free Southern Ocean into the ice field.

\begin{figure}[h]
	\centering
	\includegraphics[width=\textwidth]{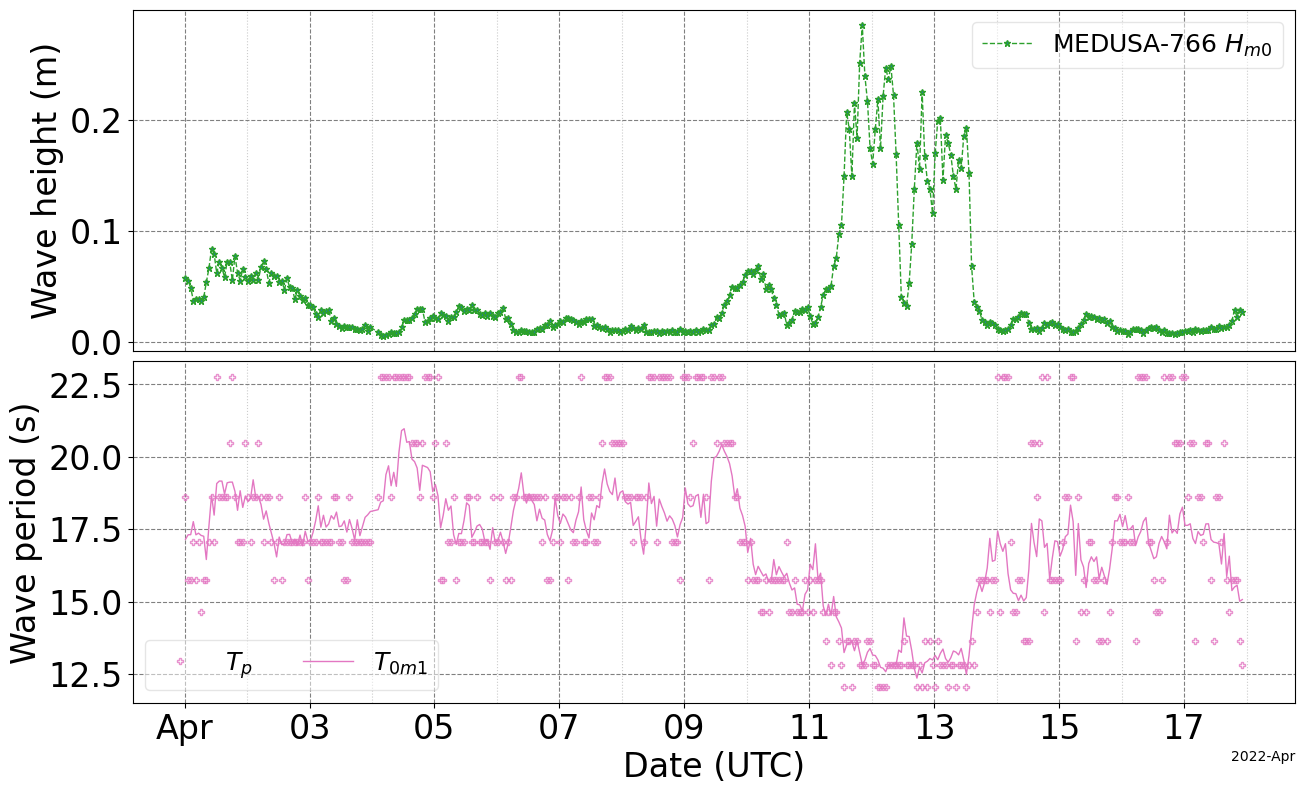}
	\caption{Medusa-766 wave height and period time series for 1--17 Apr 2022 while it was located in LBH. After 17 Apr, Medusa-766 drifted out of LBH.}
	\label{fig9:aprtimeseries}
\end{figure}

With a view to consider the long distance propagation of swell and sea ice breakup potential in the context of LHB waves, we now discuss the wave data measured while Medusa-766 was located in LHB. 
LHB is exposed to the Southern Ocean that has vast fetches and an abundance of extratropical cyclonic activities. As such, we expected frequent wave propagation to the buoy from Southern Ocean waves, at least between February and April when the SIC extent is low. However, this was not the case. When Medusa-766 was located in LHB, the measured $H_{m0}$ exceeded 0.5 m only three times in February, none in March and April. Then, from about 14 Apr, Medusa-766 began to drift westward and out of LHB; i.e., the ice floe the Medusa-766 was deployed somehow drifted out of the bay. 
We examine two synoptic events on 1 and 11 Apr 2022 before the Medusa-766 drifted out of LHB. These events generated waves towards Medusa-766 with $H_{m0}$ greater than 5 m and $T_p$ around 10--12 s at the LHB ice edge.
The wave height and period time series at Medusa-766 between 1 and 17 Apr are shown in Figure \ref{fig9:aprtimeseries}.

\begin{figure}[h]
	\centering
	\includegraphics[width=\textwidth]{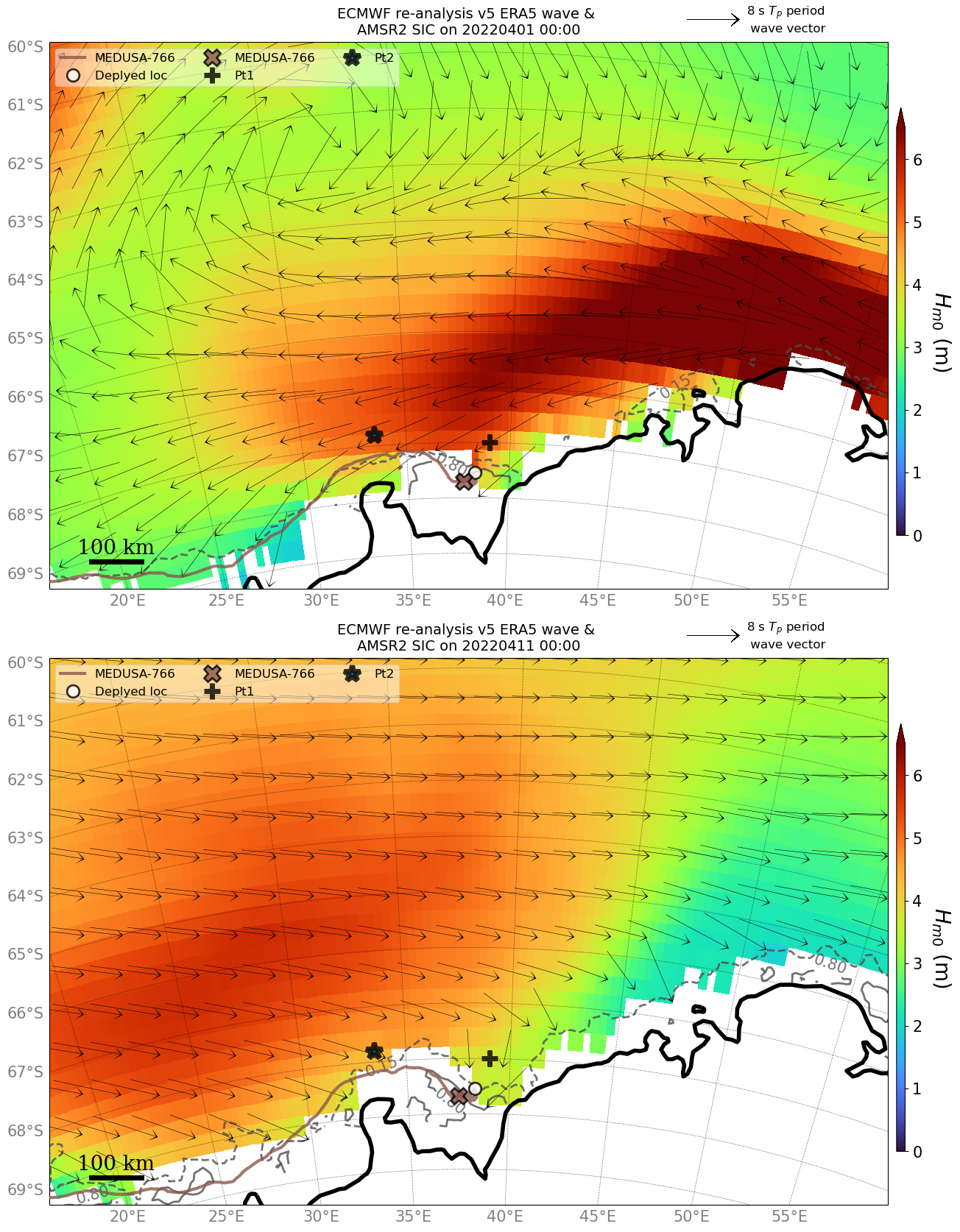}
	\caption{ERA5 wave field snapshots of two synoptic events that generated waves towards Medusa-766 near the LH bay are shown here. 1 Apr (top) was the northeast wave event when the Medusa-766 measured $H_{m0}$ was less than 0.1 m. 11 Apr was the west northwest event (bottom) when the Medusa-766 measured $H_{m0}$ was around 0.25 m. The brown line is the Medusa-766 trajectory, and the brown marker shows its location on the respective dates. The black markers show where the ERA5 directional spectra were obtained. The $H_{m0}$ is shown in the colours and the wave vectors are mean wave direction scaled by $T_p$ The 0.15 and 0.80 ADS-AMSR2 SIC contours are also shown as grey dashed and solid lines, respectively.}
	\label{fig6:LHevents}
\end{figure}

A snapshot of wave fields of the two synoptic events are presented in Figure \ref{fig6:LHevents}. The black markers, Pt1 and Pt2, indicate the estimated positions of incoming waves. The first event was a synoptic scale low pressure system located 100s of kms offshore of LHB and generated northeast waves propagating towards Medusa-766 (see the top panel of Figure \ref{fig6:LHevents}). 
At this time, Medusa-766  drifted towards ice field, then remained more or less stationary for several days. The peak of the wave event at Pt1 near the ice edge occurred at UTC 21:00 on 31 Mar with $H_{m0}$ and $T_p$ of 5.3 m and 11.2 s. The distance between Pt1 and Medusa-766 was only around 90 km, but the measured $H_{m0}$ at Medusa-766 was less than 0.1 m. 
The second event occurred via a combined effect of a low pressure system west of the Medusa-766 position and the high pressure system offshore of LHB; this generated waves travelling predominantly from west northwest as shown on the bottom panel of Figure \ref{fig6:LHevents}. The peak incoming wave energy as inferred from ERA5 was 5.1 m $H_{m0}$ and 11.2 s $T_p$. Pt2 was located around 200 km away from Medusa-766 and was more protected from the incoming waves by the ice field compared to the first event, at least as depicted by the ADS-AMSR2 SIC fields. Despite this, the Medusa-766 measured $H_{m0}$ during this event was around 0.25 m, which was larger than the first event, but still significantly attenuated. 
While the incoming wave conditions were not so dissimilar to the wave propagation in winter described in the results and the \citet{Higashi1982} event, these waves seemingly did not cause a regional scale fast ice breakup. 

We note that Medusa-766 drifted during the 11 Apr event. Three days later, on 14 Apr, Medusa-766 began to outflow and drifted westward from LBH. 
Whether ocean waves played any part in triggering the outflow of ice floes at these times is unknown, largely because the analysis and discussion in this study are based on the single buoy measurements. Building on the success of Medusa-766, 23 Medusa-OMBs were deployed in LBH (15 on fast ice and 8 in drift ice floes) during the JARE64 campaign.

\section{Conclusions} \label{conclusion}
Medusa-766 demonstrated its durability and robustness by surviving 333 days in the extreme Antarctica environment. Further, being able to detect centimetre order swell signals demonstrated the high sensitivity of the sensor used. These led to a striking illustration of the long distance propagation of swell into sea ice that has not been previously observed to the authors' knowledge, and showed the potential of the IMU-based wave measurement in sea ice. The Medusa-766 observation provided insights into the future JARE wave-ice interaction observation strategies as we aim to elucidate the relative contribution of ocean waves to the unstable LHB fast ice. 

\section*{Acknowledgement}
We are grateful to the crew and expedition members onboard the icebreaker \textit{Shirase} for their cooperative support in conducting on-deck operations during our JARE63 wave-ice buoy deployment.

\section*{Data availability}
The data that support the findings of this study are available from the corresponding author upon reasonable request.

\section*{Funding}
This work was a part of the Arctic Challenge for Sustainability II (ArCS II) Project (Program Grant Number JPMXD1420318865). \\
A part of this study was also conducted under JSPS KAKENHI Grant Numbers JP 19H00801, 19H05512, 21K14357, and 22H00241.\\

\clearpage

\section*{Table}
\begin{table}[h]
	\caption{A summary of wave attenuation coefficients and wave statistics for the 21 July event. Typical $a$ values from previous observations were obtained from table 1 of \citet{Thomson2021}, and we tuned the $n$ values to match the Medusa-766 $H_{m0}$. The incoming PSD, Medusa-766 PSD, and attenuated incoming wave PSDs over 1,250 km are shown in Figure \ref{fig7:spectra_event}. The incoming wave spectrum was obtained from ERA5 at 59$^\circ$ S, 22$^\circ$ W.}
	\label{tab:attenuation_summary}
	\begin{tabular}{|l|l|l|l|l|l|}
		\hline
		\textbf{PSD label}    & \textbf{$a$} & \textbf{$n$} & \textbf{$H_{m0}$ (m)} & \textbf{$T_p$ (s)} & \textbf{$T_{0m1}$ (s)} \\ \hline \hline
		Incoming PSD $S_{in}$ & -          & -          & 3.250                 & 14.86              & 13.40                  \\ \hline
		Medusa-766 $S_{ice}$ & -          & -          & 0.075                 & 18.62              & 18.48                  \\ \hline
		Attenuated PSD $S_{in}$0 & 0.005      & 2.41       & 0.073                 & 18.00              & 18.53                  \\ \hline
		Attenuated PSD $S_{in}$1 & 0.010      & 2.65       & 0.074                 & 18.00              & 18.75                  \\ \hline
		Attenuated PSD $S_{in}$2 & 0.020      & 2.89       & 0.075                 & 19.78              & 18.95                  \\ \hline
		Attenuated PSD $S_{in}$3 & 0.025      & 2.97       & 0.077                 & 19.78              & 19.00                  \\ \hline
	\end{tabular}
\end{table}

\section*{Figures}

\clearpage

\bibliographystyle{tfcad}
\bibliography{ref}

\begin{thebibliography}{30}
\newcommand{\enquote}[1]{``#1''}
\providecommand{\natexlab}[1]{#1}
\providecommand{\url}[1]{\normalfont{#1}}
\providecommand{\urlprefix}{}

\bibitem[Ahn(2021)]{Ahn2021}
Ahn, Seongho. 2021. ``Modeling mean relation between peak period and energy
  period of ocean surface wave systems.'' \emph{Ocean Engineering} 228: 108937.

\bibitem[Boutin et~al.(2018)]{Boutin2018}
Boutin, Guillaume, Fabrice Ardhuin, Dany Dumont, Caroline Sévigny, Fanny
  Girard-Ardhuin, and Mickael Accensi. 2018. ``Floe Size Effect on Wave-Ice
  Interactions: Possible Effects, Implementation in Wave Model, and
  Evaluation.'' \emph{Journal of Geophysical Research: Oceans} 123 (7):
  4779--4805.

\bibitem[Collins and Jensen(2022)]{Collins2022}
Collins, C.~O., and R.~E. Jensen. 2022. ``Tilt Error in NDBC Ocean Wave Height
  Records.'' \emph{Journal of Atmospheric and Oceanic Technology} 39 (7).

\bibitem[Dumont, Kohout, and Bertino(2011)]{Dumont2011}
Dumont, D., A.~Kohout, and L.~Bertino. 2011. ``A wave-based model for the
  marginal ice zone including a floe breaking parameterization.'' \emph{Journal
  of Geophysical Research: Oceans} 116 (C4).

\bibitem[Higashi et~al.(1982)]{Higashi1982}
Higashi, Akira, Dougal~J. Goodman, Sadao Kawaguchi, and Shinji Mae. 1982. ``The
  cause of the breakup of fast ice on March 18, 1980 near Syowa Station, East
  Antarctica.'' \emph{Memoirs of National Institute of Polar Research. Special
  issue} 24: 222--231.

\bibitem[Hori et~al.(2012)]{ADS-AMSR2}
Hori, M., H.~Yabuki, T.~Sugimura, and T.~Terui. 2012. ``{AMSR2 Level 3} product
  of Daily Polar Brightness Temperatures and Product, 1.00.'' [Accessed on
  2023/03/21],  \urlprefix\url{https://ads.nipr.ac.jp/dataset/A20170123-003}.

\bibitem[Karulina et~al.(2019)]{Karulina2019}
Karulina, M., A.~Marchenko, E.~Karulin, D.~Sodhi, A.~Sakharov, and
  P.~Chistyakov. 2019. ``Full-scale flexural strength of sea ice and freshwater
  ice in Spitsbergen Fjords and North-West Barents Sea.'' \emph{Applied Ocean
  Research} 90: 101853.

\bibitem[Kohout et~al.(2015)]{Kohout2015}
Kohout, Alison~L., Bill Penrose, Scott Penrose, and Michael~J.M. Williams.
  2015. ``A device for measuring wave-induced motion of ice floes in the
  {Antarctic} Marginal Ice Zone.'' \emph{Annals of Glaciology} 56 (69):
  415--424.

\bibitem[Liu and Mollo-Christensen(1988)]{Liu1988}
Liu, Antony~K., and Erik Mollo-Christensen. 1988. ``Wave Propagation in a Solid
  Ice Pack.'' \emph{Journal of Physical Oceanography} 18 (11): 1702 -- 1712.

\bibitem[Meylan et~al.(2018)]{Meylan2018}
Meylan, M.~H., L.~G. Bennetts, J.~E.~M. Mosig, W.~E. Rogers, M.~J. Doble, and
  M.~A. Peter. 2018. ``Dispersion Relations, Power Laws, and Energy Loss for
  Waves in the Marginal Ice Zone.'' \emph{Journal of Geophysical Research:
  Oceans} 123 (5): 3322--3335.

\bibitem[Moreau, Weiss, and Marsan(2020)]{Moreau2020}
Moreau, Ludovic, Jérôme Weiss, and David Marsan. 2020. ``Accurate Estimations
  of Sea-Ice Thickness and Elastic Properties From Seismic Noise Recorded With
  a Minimal Number of Geophones: From Thin Landfast Ice to Thick Pack Ice.''
  \emph{Journal of Geophysical Research: Oceans} 125 (11): e2020JC016492.

\bibitem[Nose et~al.(2023)]{Nose2023}
Nose, Takehiko, Jean Rabault, Takuji Waseda, Tsubasa Kodaira, Yasushi Fujiwara,
  Tomotaka Katsuno, Naoya Kanna, Kazutaka Tateyama, Joey Voermans, and Tatiana
  Alekseeva. 2023. ``A comparison of an operational wave–ice model product
  and drifting wave buoy observation in the central Arctic Ocean: investigating
  the effect of sea-ice forcing in thin ice cover.'' \emph{Polar Research} 42.

\bibitem[Nose et~al.(2018)]{Nose2018}
Nose, Takehiko, Adrean Webb, Takuji Waseda, Jun Inoue, and Kazutoshi Sato.
  2018. ``Predictability of storm wave heights in the ice-free Beaufort Sea.''
  \emph{Ocean Dynamics} 68 (10): 1383--1402.

\bibitem[Rabault(2022)]{OMBGit}
Rabault, Jean. 2022. ``{OpenMetBuoy-v2021a}.''
  \url{https://github.com/jerabaul29/OpenMetBuoy-v2021a/tree/main/legacy_firmware/firmware/two_ways_gps_waves_drifter}.
  [The OMB Github repository. Broken hyperlink].

\bibitem[Rabault et~al.(2023)]{Rabault2023}
Rabault, Jean, Malte Muller, Joey Voermans, Dmitry Brazhinov, Ian Turnbull,
  Aleksey Marchenko, Martin Biuw, et~al. 2023. ``A dataset of direct
  observations of sea ice drift and waves in ice.'' \emph{Scientific Data} 10
  (251).

\bibitem[Rabault et~al.(2022)]{Rabault2022}
Rabault, Jean, Takehiko Nose, Gaute Hope, Malte M\"{u}ller, {\O}yvind Breivik,
  Joey Voermans, Lars~Robert Hole, et~al. 2022. ``{OpenMetBuoy}-v2021: An
  Easy-to-Build, Affordable, Customizable, Open-Source Instrument for
  Oceanographic Measurements of Drift and Waves in Sea Ice and the Open
  Ocean.'' \emph{Geosciences} 12 (3): 110.
  \urlprefix\url{https://doi.org/10.3390/geosciences12030110}.

\bibitem[Rabault et~al.(2020)]{Rabault2020}
Rabault, Jean, Graig Sutherland, Olav Gundersen, Atle Jensen, Aleksey
  Marchenko, and Øyvind Breivik. 2020. ``An open source, versatile, affordable
  waves in ice instrument for scientific measurements in the Polar Regions.''
  \emph{Cold Regions Science and Technology} 170: 102955.
  \urlprefix\url{https://www.sciencedirect.com/science/article/pii/S0165232X19300230}.

\bibitem[RockBLOCK(2023)]{Rockblock}
RockBLOCK. 2023. ``Power Consumption Guidance.''
  \url{https://docs.rockblock.rock7.com/docs/power-consumption-guidance#:~:text=The%20current%20consumption%20in%20Idle,Ring%20Alerts%20can%20be%20received.}
  (Accessed on 2023/03/18. Broken hyperlink).

\bibitem[STMicroelectronics(2020)]{STMicro}
STMicroelectronics. 2020. ``Datasheet - ISM330DHCX - iNEMO inertial module with
  embedded Machine Learning Core: always-on 3D accelerometer and 3D gyroscope
  with digital output for industrial applications.''
  \url{https://www.st.com/resource/en/datasheet/ism330dhcx.pdf}, November.
  DS13012 - Rev 7. (Accessed on 2023/03/23.).

\bibitem[Thomson et~al.(2021)]{Thomson2021}
Thomson, Jim, Lucia Hošeková, Michael~H. Meylan, Alison~L Kohout, and
  Nirnimesh Kumar. 2021. ``Spurious Rollover of Wave Attenuation Rates in Sea
  Ice Caused by Noise in Field Measurements.'' \emph{Journal of Geophysical
  Research: Oceans} 126 (3): e2020JC016606.

\bibitem[Timco and Weeks(2010)]{Timco2010}
Timco, G.W., and W.F. Weeks. 2010. ``A review of the engineering properties of
  sea ice.'' \emph{Cold Regions Science and Technology} 60 (2): 107--129.

\bibitem[Ushio(2003)]{Ushio2003}
Ushio, Shuki. 2003. ``Frequent sea-ice breakup in Lutzow-Holmbukta, Antarctica,
  based on analysis of ice condition from 1980 to 2003.'' \emph{Antarctic
  Record} 47 (3): 338--348. [In Japanese with English abstract].

\bibitem[Ushio(2006)]{Ushio2006}
Ushio, Shuki. 2006. ``Factors affecting fast-ice break-up frequency in
  Lützow-Holm Bay, Antarctica.'' \emph{Annals of Glaciology} 44: 177–182.

\bibitem[Ushio et~al.(2004)]{Ushio2004}
Ushio, Shuki, Shotaro Uto, Koh Izumiyama, Haruhito Shimoda, and Masaru Ayukawa.
  2004. ``Interannual variation of landfast ice condition in Lutzow-Holmbukta,
  Antarctica, derived from navigation log of icebreaker Shirase.''
  \emph{Antarctic Record} 48 (3): 180--190. [In Japanese with English
  abstract].

\bibitem[Voermans et~al.(2021)]{Voermans2021}
Voermans, J.~J., Q.~Liu, A.~Marchenko, J.~Rabault, K.~Filchuk, I.~Ryzhov,
  P.~Heil, et~al. 2021. ``Wave dispersion and dissipation in landfast ice:
  comparison of observations against models.'' \emph{The Cryosphere} 15 (12):
  5557--5575.
  \urlprefix\url{https://tc.copernicus.org/articles/15/5557/2021/}.

\bibitem[Voermans et~al.(2020)]{Voermans2020}
Voermans, J.~J., J.~Rabault, K.~Filchuk, I.~Ryzhov, P.~Heil, A.~Marchenko,
  C.~O. Collins~III, M.~Dabboor, G.~Sutherland, and A.~V. Babanin. 2020.
  ``Experimental evidence for a universal threshold characterizing wave-induced
  sea ice break-up.'' \emph{The Cryosphere} 14 (11): 4265--4278.
  \urlprefix\url{https://tc.copernicus.org/articles/14/4265/2020/}.

\bibitem[Voermans et~al.(2023)]{Voermans2023}
Voermans, Joey~J., Jean Rabault, Aleksey Marchenko, Takehiko Nose, Takuji
  Waseda, and Alexander~V. Babanin. 2023. ``Estimating the elastic modulus of
  landfast ice from wave observations.'' \emph{Journal of Glaciology} 1–11.

\bibitem[Wadhams(1975)]{Wadhams1975}
Wadhams, Peter. 1975. ``Airborne laser profiling of swell in an open ice
  field.'' \emph{Journal of Geophysical Research (1896-1977)} 80 (33):
  4520--4528.

\bibitem[Waseda et~al.(2022)]{Waseda2022}
Waseda, Takuji, Alberto Alberello, Takehiko Nose, Takenobu Toyota, Tsubasa
  Kodaira, and Yasushi Fujiwara. 2022. ``Observation of anomalous spectral
  downshifting of waves in the Okhotsk Sea Marginal Ice Zone.''
  \emph{Philosophical Transactions of the Royal Society A: Mathematical,
  Physical and Engineering Sciences} 380 (2235): 20210256.

\bibitem[Waseda et~al.(2017)]{Waseda2017}
Waseda, Takuji, Adrean Webb, Kazutoshi Sato, and Jun Inoue. 2017. ``Arctic Wave
  Observation by Drifting Type Wave Buoys in 2016.'' In \emph{The 27th
  International Ocean and Polar Engineering Conference}, San Francisco,
  California, USA. International Society of Offshore and Polar Engineers.

\end{thebibliography}

\clearpage
\appendix
\section{Iridium transmission performance and power consumption estimates}\label{Iridium_analysis}
\subsection{Iridium transmission}
Sending measured data via the Iridium messaging is one of the most important and difficult logic to control. This is because the number of factors that affect the successful transmission is largely out of our control and sometimes unpredictable. For example, even when we test in what seems an ideal environment (e.g., an open area at the University of Tokyo Kashiwa campus), Iridium messaging sometimes fails. The unpredictability is exacerbated in the harsh polar environments.

We evaluated the GNSS and wave data transmission and found that the data capture rates were approximately 95 \% for both GNSS and wave measurements; these capture rates reflect the robustness of the OMB design, e.g., the last-in first-out buffering of data packets and the Iridium transmission retry strategy implemented in the firmware. Nevertheless, we examined how readings were skipped so that the OMB firmware logic may be further improved. The majority of the data losses occurred when the Iridium transmission stopped for prolonged period of time up to five days, and the MCU somehow reboots and looses stored data in the buffer. The reboot could be identified when the sampling intervals reverted to the default values and/or GNSS or wave readings were made at irregular times, i.e., not at minute 0 or 30 of each hour. From the record, we identified that Medusa-766 rebooted at least 16 times, of which 5 of them occurred after 20 Dec 2022 when the Iridium messages became irregular. 

\begin{figure}[h]
	\centering
	\includegraphics[width=\textwidth]{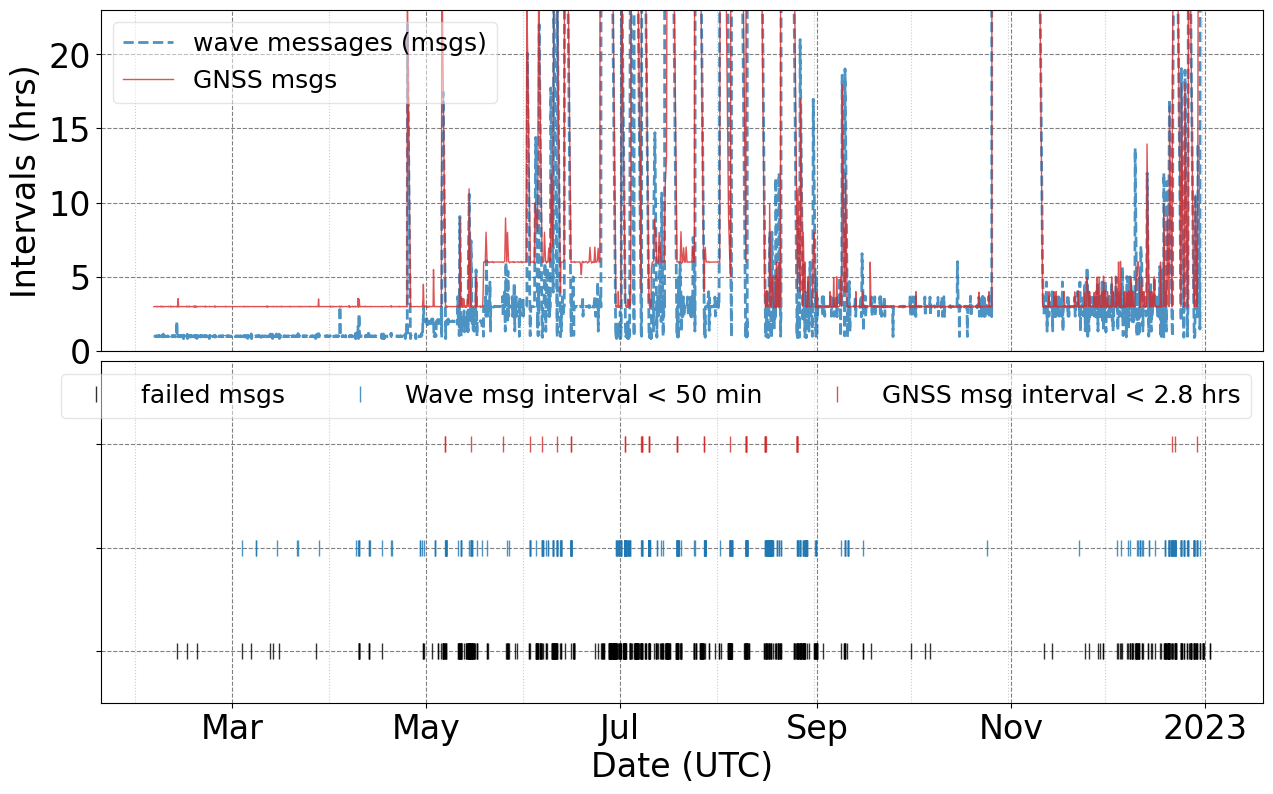}
	\caption{Iridium transmission intervals for both GNSS (red) and wave (blue) messages (top). The failed transmission are shown on the bottom panel. The red and blue markers show when the GNSS and wave messages were sent at shorter intervals than the default intervals, which indicates the buffered messages that were stored due to failed transmissions were being transmitted. Despite the seemingly many failed transmission, the data capture rate during the observation was \textasciitilde95 \%.}
	\label{fig1:Iridium_log}
\end{figure}

The most critical error regarding wave measurements occurred when the wave readings were missed between 12:00 5 Jul and 09:00 7 Jul 2022 while the GNSS readings were measured correctly. It remains unknown how this error was caused, but the QWIIC connection (\url{https://www.sparkfun.com/qwiic}) between the MCU and the IMU may become unstable during thermal cycling. Direct soldering between the MCU and the IMU is a future OMB improvement. Another recurring cause of missed readings seemed to occur when there were consecutive GNSS readings taken at minute 0 of an hour, which then causes the next GNSS and wave readings to be missed. This error occurred 24 times during the deployment period. Interestingly, these consecutive GNSS readings only caused the next readings to be missed and did not cause the MCU to reboot. Lastly, there were 2 and 9 times when there was isolated missed readings for GNSS and wave data, respectively. Regardless of the error types, these errors seemed to occur more often when there was more difficulty with Iridium transmissions as reflected by the number of failed Iridium transmission attempts shown in the bottom panel of Figure \ref{fig1:Iridium_log}. From late November and December, the Iridium messages became noticeably irregular, which coincides with the time when the sea ice rapidly melted in the observation region. 

\subsection{Power consumption estimates}
The Li primary battery is the choice of batteries at low temperature environments. With that comes the difficulty of monitoring/estimating battery life because the Li primary batteries have a flat discharge curve for most of its life. Here, we attempt to visualise the power consumption estimates for the major OMB tasks during the nearly year-long Medusa-766 measurement duration. This is shown as the consumed battery capacity as calculated by $C=ct$ where $C$ is the battery capacity in Amp-hour, $c$ is the current, and $t$ is the time. Their respective estimated time of use are 2 and 20 minutes for GNSS and IMU sensing; the MCU was considered to be in the sleep mode for the remainder of the time. 
The current estimated in \citet{Rabault2022} and Medusa-766 differed because being a prototype, we did not disable the user-friendly functioning indicator LEDs. During a different measurement preparation in 2022, we conducted a power consumption test to compare the current with and without LEDs and found there are considerable differences in power consumption. The three primary modes of OMB operations are GNSS and IMU sensing, and sleep. The currents used for each were as follows (with/without LEDs): GNSS sensing; 45/30 mA, IMU sensing; 15/5 mA, and sleep; 7/0.5 mA. 

The Iridium module power consumption is the trickiest part of this exercise as the maximum required current draw is large and in short bursts; it is therefore not practical to measure the currents with a multi meter. Here, we rely on the Iridium module manufacturer's power consumption guidance \citep{Rockblock}. The relevant functions and its required battery capacity are summarised here. The battery capacities used to produce the Iridium module power consumption are underlined and in the bold text. 
\begin{itemize}
	\item \textbf{charging mode:} when the device is first powered or after a long time of disuse (typically more than a week), the Iridium module goes in the charging mode to charge the supercapacitor that acts as an energy reservoir and serves to buffer the highly pulsed nature of its internal circuits from the user connections. The battery capacity needed for the charging phase is given in \citet{Rockblock} as \underline{\textbf{12.5 Amp-s}}.
	\item \textbf{active mode:} this is the transmitting or other command driven activity. Battery capacity required for the successful transmission is estimated to be \underline{\textbf{2.8 Amp-s}}, and the failed transmission consumes slightly more power at \underline{\textbf{3.8 Amp-s}}. These numbers assume the average time to complete a transmission (exit the sleep mode, initialise the unit, send a message, and for the current to drop back to its prior level) is one minute.  
	\item \textbf{sleep mode:} when the circuit is turned off, the Iridium module uses 73 $\mu$A. However, it takes 24 hours to drop to this true sleep current level, so the effective sleep current for less than 24 hours of disuse is given as 0.1 mA, which is typically reached after an hour. So the battery capacity needed for an hour of sleep is \underline{\textbf{0.36 Amp-s}}.
	\item \textbf{active-sleep mode transition:} the transition from the active to sleep modes takes around one hour and has been measured to be approximately \underline{\textbf{4 Amp-s}}. In other words, when the Iridium transmission intervals are less than 1 hour, we use this transition battery capacity as the Iridium module has no time to reach its sleep mode current level.
\end{itemize}

Figure \ref{fig1:Iridium_log} summarised the Iridium transmission records. The top panel shows the transmission intervals for GNSS and wave messages. The black markers on the bottom panel shows the times failed Iridium transmissions occurred as reflected by an empty payload. Note that there are likely failed transmissions that are not sent to the RockBLOCK server as empty payload; therefore this remains a source of uncertainty. The red and blue markers on the bottom panel show when the GNSS and wave messages were sent at shorter intervals than the default intervals, which indicate the times when the Iridium messaging was catching up, i.e.,  failed messages stored in the buffer were being transmitted. It is clear from this figure that unfavourable conditions for Iridium transmissions occurred more frequently in May, June, July, August, and Dec. Note that there was an approximate 15 day period of missed readings between 26 Oct and 10 Nov 2022 because we let the Iridium line rental expire and forgot to initially renew it. From communication with RockBLOCK, the device functions as normal (i.e., the Iridium modem does not need to be active to complete the process of successful transmission), but the data are discarded and not received by the RockBLOCK server.	(Pers. Comms. \url{iridium.support@groundcontrol.com} on 27 Apr 2022).

\begin{figure}[h]
	\centering
	\includegraphics[width=\textwidth]{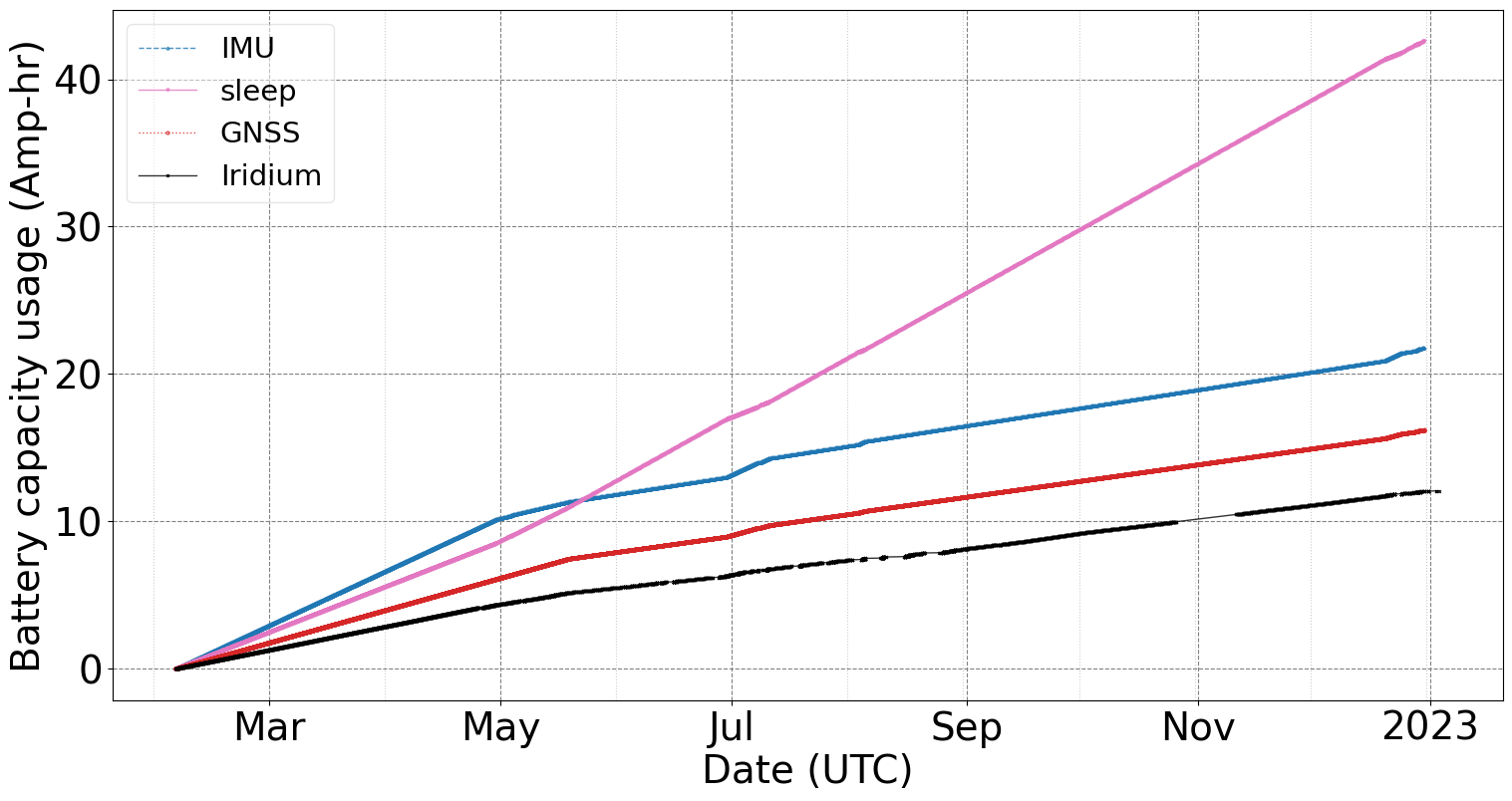}
	\caption{Battery capacity usage estimates for GNSS (red) and IMU (blue) sensing, MCU sleep (pink), and Iridium transmission (black) during the Medusa-766 measurement period.}
	\label{fig2:power_consumption}
\end{figure}

Using these transmission data, the Iridium battery capacity usage was estimated and shown in Figure \ref{fig2:power_consumption}. To be conservative, we assumed that the 12.5 Amp-hr was consumed to charge the supercapacitor when the device was offline for more than 24 hours. The total estimated battery capacity for GNSS and IMU sensing, MCU sleep, and the Iridium module were 16.2, 21.8, 42.6, and 12.1 Amp-hr, respectively. The most striking outcome is that the MCU sleep consumed nearly twice as much as any of the primary functions. Considering that the MCU sleep uses $\frac{1}{10}$th of the IMU breakout when the LEDs are disabled, compared with that of nearly $\frac{1}{2}$ with the LEDs on, it is clear that disabling the LEDs is a critical step in the OMB building procedure for efficient utilisation of the OMB (as per the OMB assembly instructions \citep{OMBGit}). Another interesting and encouraging results here is that Iridium transmission was the least power consumed task, which is contrary to that estimated in \citet{Rabault2022}. 

Despite not disabling the LEDs, the total battery capacity consumed was estimated to be 92.7 Amp-hr, so the remaining capacity was 21.3 Amp-hr, roughly equivalent to the single Tadiran TL-5930 battery capacity. This is encouraging as it appears we can estimate the battery life to $\pm$ one battery capacity even when the sensor endured the extreme low temperature of Antarctica winter.

\end{document}